\documentclass[10pt,a4paper,superscriptaddress,aps,prd,nofootinbib,notitlepage]{revtex4-1}
\usepackage{lmodern}

\usepackage[T1]{fontenc}
\usepackage[utf8]{inputenc}
\usepackage{color}
\usepackage{amsmath}
\usepackage{amssymb}
\usepackage{esint}
\usepackage[unicode=true,
 bookmarks=true,bookmarksnumbered=false,bookmarksopen=false,
 breaklinks=false,pdfborder={0 0 1},backref=false,colorlinks=true]
 {hyperref}
\hypersetup{pdftitle={GW speed: Implications for models without mass scale},
 pdfauthor={Henrik Nersisyan, Nelson Lima, Luca Amendola},
 pdfkeywords={dark energy, modified gravity, large-scale structure},
 citecolor=red,filecolor=blue,linkcolor=blue,urlcolor=blue}
\usepackage{breakurl}



\begin{document}

\title{Gravitational wave speed: Implications for models without a mass
scale}

\author{Henrik Nersisyan, Nelson A. Lima and Luca Amendola}

\affiliation{ITP, Ruprecht-Karls-Universität Heidelberg, Philosophenweg 16, 69120
Heidelberg, Germany}
\begin{abstract}
The recent report that the gravitational wave speed equals the light
speed puts strong constraints on the anisotropic stress parameter
of many modified gravity models, a quantity that is directly observable
through large-scale structure. We show here that models without a
mass scale completely escape these constraints. We discuss a few relevant
cases in detail: Brans-Dicke theory, nonlocal models, and Galileon
Lagrangian. 
\end{abstract}

\maketitle

\section{Introduction}

The recent detection of gravitational waves (GW) from a neutron star
merger~\cite{TheLIGOScientific:2017qsa} had a big impact in the
field of modified gravity theories~\cite{Lombriser:2016,Amendola:2017orw,Creminelli:2017sry,Ezquiaga:2017ekz,Sakstein:2017xjx,Baker:2017hug}.
For the first time we observed a multi-messenger event, with a simultaneous
detection of gravitational waves and an associated optical counterpart.
This observation placed a constraint on the speed of gravitational
waves $c_{\text{T}}$ with an unprecedented accuracy of $\vert c_{\text{T}}/c-1\vert\leq1\times10^{-15}$~\cite{Monitor:2017mdv},
where $c$ is the speed of light. Therefore, at least at the present
time, gravitational waves undoubtedly travel at the speed of light.

The GW speed constitutes a smoking gun evidence for several modified
gravity theories where modifications in the gravitational sector affect
the propagation of the gravitational degrees of freedom. Many of these
models fall within the so-called Horndeski gravity, which is the most
general four-dimensional scalar-tensor theory with second order equations
of motion \cite{Horndeski:1974} (although recently Horndeski gravity
has been generalized to include higher-order derivative terms while
keeping only one additional scalar degree of freedom~\cite{Zumalacarregui:2013pma,Gleyzes:2014dya,Deffayet:2015qwa,Crisostomi:2016tcp}).
Some of the terms present in the Horndeski action lead to a modification
of gravitational wave speed which puts them, and the theories they
define, in great tension with the aforementioned constraint on $c_{\text{T}}$~\cite{Ezquiaga:2017ekz}.
In passing, let us notice that the constraints on modified gravity
models disappear if in the frame in which baryons are uncoupled the
gravity sector is standard; this leaves completely free the coupling
between dark matter and dark energy. Here, however, we restrict our
attenton to universally-coupled models.

Modifications of gravitational wave propagation are also linked to
the large scale structure formation in modified gravity theories,
as first shown in Refs.~\cite{Saltas:2014dha,Sawicki:2016klv}. Hence,
by constraining the speed of gravitational waves we can also constrain
$\eta$. The latter, for scalar perturbations in the Newtonian gauge
with line element 
\begin{equation}
ds^{2}=-\left(1+2\Psi\right)dt^{2}+a^{2}\left(1+2\Phi\right)\delta_{ij}dx^{i}dx^{j},
\end{equation}
where $a(t)$ is the cosmological scale factor, is defined as the
ratio of Bardeen potentials $\Phi$ and $\Psi$, i.e., 
\begin{equation}
\eta\equiv-\frac{\Phi}{\Psi}.
\end{equation}
Furthermore, one can also define the effective Gravitational coupling
$Y$, in momentum space, as 
\begin{equation}
Y\equiv-\frac{2k^{2}\Psi}{a^{2}\rho_{\text{m}}\delta_{\text{m}}},
\end{equation}
where $\rho_{\text{m}}$ and $\delta_{\text{m}}$ are the matter energy
density and the matter density contrast, respectively\footnote{All the perturbation quantities in this paper are meant to denote
the positive-definite root-mean-squares of the corresponding random
variables.}.

In Refs.~\cite{Saltas:2014dha,Sawicki:2016klv} it was shown that
in the quasi-static approximation $\left(\hat{k}\equiv k/Ha\gg1\right)$,
for scales much smaller than the mass $M$ of scalar field fluctuations
$k\ll M$, the gravitational slip parameter takes the value $\eta=c_{\text{T}}^{-2}$.
Hence, by using the fact that gravitational waves, at least presently,
travel with the speed of light, one obtains that $\eta$ should also
be equal to unity at those scales~\cite{Amendola:2017orw}, i.e.
it should recover its General Relativity (GR) value. However, in this
work we want to develop the suggestion advanced in \cite{Amendola:2017orw}
that this constraint on $\eta$ is applicable only for those modifications
of gravity which introduce a new mass scale into the theory: when
a theory of gravity is not supplemented with a new mass scale, $\eta$
is \emph{a}) scale-independent, \emph{b}) in general different from
unity, and \emph{c}) left unconstrained by the GW constraint $c_{\text{T}}=c$.

In the case of Horndeski theory, in the quasi-static approximation
and inside the Jeans length of the scalar field $\left(kc_{\text{s}}\gg1\right)$,
we have for the gravitational slip parameter $\eta$ and the effective
gravitational constant $Y$ the following expressions~\cite{Amendola:2012ky,DeFelice:2011bh,DeFelice:2011hq}
\begin{align}
\eta & =h_{2}\left(\frac{1+k^{2}h_{4}}{1+k^{2}h_{5}}\right),\hspace{10mm}Y=h_{1}\left(\frac{1+k^{2}h_{5}}{1+k^{2}h_{3}}\right).\label{eq:etay}
\end{align}
Here, the functions $h_{1-5}$ express the modification introduced
by Horndeski gravity compared to GR, and are scale independent functions
constructed from the Horndeski functional parameters and their derivatives
with respect to the scalar field $\phi$ and its canonical kinetic
term $X=-g_{\mu\nu}\phi^{,\mu}\phi^{,\nu}/2$, i.e. $h_{i}\equiv h_{i}\left(z\right)\equiv h_{i}\left(\phi,X\right)$.
One can find explicit expressions for the Horndenski parameters $h_{i}$
in the Appendix of Ref.~\cite{Amendola:2012ky}. It turns out that
the Horndeski function $h_{2}$ in this approximation has a simple
relation with the speed of gravitational waves, namely $h_{2}=c_{\text{T}}^{-2}$.
For the case of $\Lambda\text{CDM}$ cosmology we have that $h_{1,2}=1$
and $h_{3,4,5}=0$, hence $\eta$ is identically equal to unity.

However, from Eq.~(\ref{eq:etay}) it should be clear that in the
limit $k^{2}h_{4},k^{2}h_{5}\gg1$, it is possible to have $\eta=h_{2}h_{4}/h_{5}$.
This implies a value of the slip that is generally different from
unity and currently unconstrained by the speed of gravitational waves.
As we will argue forward for theories which do not introduce any additional
mass scale besides the Planck mass, a similar conclusion holds at
all the scales within the quasi-static approximation.

\section{A prototypical example: Brans-Dicke model}

First, we will focus on one of the most well-known examples of modified
theories of gravity, Brans-Dicke~\cite{Brans:1961sx}. In the Jordan
frame, its action can be written as 
\begin{equation}
S_{{\rm {BD}}}=\frac{1}{16\pi G}\int d^{4}x\sqrt{-g}\Big[\phi R-\frac{\omega_{{\rm {BD}}}}{\phi}(\partial\phi)^{2}-V(\phi)\Big]+S_{{\rm {m}}},\label{bdaction}
\end{equation}
where $S_{{\rm {m}}}$ corresponds to the matter action, $\omega_{{\rm {BD}}}$
is the Brans-Dicke parameter and $V(\phi)$ is the self-interacting
potential that can be taken, for instance, as a cosmological constant.
From Eq.~(\ref{bdaction}) it should be clear that the dimensionless
scalar field $\phi$ performs an effective rescaling of the gravitational
constant. The General Relativity limit of the theory is thus recovered
when $\omega_{{\rm {BD}}}\rightarrow\infty$ and $\phi\rightarrow1$.
Currently, the tightest constraints available on the Brans-Dicke parameter
come from Solar-System tests, placing a lower bound of $40000$ on
$\omega_{{\rm {BD}}}$~\cite{clifford:2006,cassini:2003}. On cosmological
scales, this bound is relaxed by approximately one order of magnitude~\cite{avilez:2013,planckbd:2016}
but, on the other hand, it applies also to the case in which baryons
are uncoupled.

Despite its apparent simplicity, Brans-Dicke introduces an interesting
phenomenology across all periods of cosmic history. At early times,
it is well established that the BD scalar field follows distinct attractor
solutions: it remains frozen during radiation domination and follows
a power-law of the scale factor $a$ during matter domination. Given
the role of $\phi$, the rescaling of the cosmological gravitational
constant $G_{{\rm {cos}}}=G_{{\rm {N}}}/\phi$ produces a shift in
the background expansion history compared to $\Lambda$CDM, that is
suppressed as we approach $z=0$ if one fixes the present-day value
of the scalar field to $\phi_{0}\approx1$.

The modifications introduced by the Brans-Dicke theory also extend
to the linear evolution of the gravitational potentials, $\Phi$ and
$\Psi$. In the regime of validity of the quasi-static approximation,
the theory predicts a non-trivial value for the slip between the gravitational
potentials given by~\cite{DeFelice:2011hq} 
\begin{equation}
\eta=\frac{1+\omega_{{\rm {BD}}}+\phi(Ma/k)^{2}}{2+\omega_{{\rm {BD}}}+\phi(Ma/k)^{2}},\label{bd_eta}
\end{equation}
where $M^{2}\approx V_{\phi\phi}$ is the mass scale introduced by
the Brans-Dicke theory. It is evident that if $M=0$ as, for instance,
for the case of a constant potential $V(\phi)$, the mass scale is
removed and $\eta$ becomes scale independent. If we take into account
the cosmological constraints on the Brans-Dicke parameter, roughly
$\omega_{{\rm {BD}}}<1000$ \cite{avilez:2013,planckbd:2016}, $\eta$
could still deviate from unity at the $10^{-3}$ level.

\section{Nonlocal models}

As another less trivial example, we choose the Deser-Woodard (DW)
nonlocal gravity model where the new terms added to the Einstein-Hilbert
action do not introduce any additional dimension scale into the theory~\cite{Deser:2007jk}.
Indeed the action of the DW model is given by 
\begin{equation}
S_{\text{DW}}=\frac{1}{16\pi G}\int\sqrt{-g}d^{4}x\left[R+Rf\left(\frac{R}{\Box}\right)\right]+S_{\text{m}}.\label{dwaction}
\end{equation}
In this action the modification of gravity is achieved by the inclusion
of the term $f\left(R/\Box\right)$. The function $f$ is a general
analytic function and should be chosen to reproduce a cosmologically
valid evolution of our Universe, both at the background and perturbation
levels~\cite{Woodard:2014iga,Deffayet:2009ca}. The absence of a
new scale within the DW model is explained by the fact that the Ricci
scalar $R$ and d' Alambert operator $\Box$ have the same dimension
so the argument of the function $f$ is a dimensionless quantity.
Phenomenological studies of the DW model are performed usually by
localizing it~\cite{Nersisyan:2017mgj,Park:2017zls}. This is done
by introducing two scalar fields $X$ and $U$ defined through the
following differential equations 
\begin{equation}
\Box X\equiv R,\hspace{15mm}\Box U\equiv f'R,\label{auxfields}
\end{equation}
where $f'$ stands for the derivative of the function $f$ with respect
to its argument. Using auxiliary fields $X$ and $U$ we can translate
the initial nonlocal theory~(\ref{dwaction}) into a local multi-scalar
one. Investigation of the DW model at the background level have been
performed in a way to reproduce exact $\Lambda\text{CDM}$ behavior.
This has been achieved by choosing a particular structure for the
function $f$~\cite{Deffayet:2009ca}. After fixing the structure
of the function $f$ there are no more free parameter in the model,
and we can proceed to investigate the behavior of the DW model at
the perturbation level. The perturbative studies of the DW model show
that it has a distinct behavior from $\Lambda\text{CDM}$ at this
level. Moreover, from the observational point of view it can also
be favored over $\Lambda\text{CDM}$ model~\cite{Nersisyan:2017mgj,Park:2017zls}.
The gravitational slip parameter for the DW model has the structure
\begin{equation}
\eta=-\frac{1+U+f-4f'}{1+U+f-8f'}.\label{slipDW}
\end{equation}
This expression shows that, in the quasi-static approximation, the
gravitational slip parameter does not have a scale dependence and
can have a value different from unity. In Ref. \cite{Nersisyan:2017mgj}
it was found numerically $\eta\approx-0.05+1.79a^{2}-0.54a^{3}$.
So for the DW model in the quasi-static approximation there is no
limit where we recover the GR value of $\eta$.

The scale independence of $\eta$ in the DW model can be directly
understood from the structure of the perturbation equations. Here,
for simplicity, we show just the perturbation of the equation corresponding
to the auxiliary field $X$. However, the arguments presented below
also hold for the other fields introduced in the theory. For the linear
perturbation $\delta X$ of the auxiliary field $X$ one has the following
equation 
\begin{eqnarray}
 &  & \delta\ddot{X}+3H\delta\dot{X}+\frac{k^{2}}{a^{2}}\delta X=\label{eq:perX}\\
 &  & -6H\dot{\Phi}-6\ddot{\Phi}+\left(\dot{\Psi}+3\dot{\Phi}\right)\left(\dot{X}+6H\right)-2\frac{k^{2}}{a^{2}}\left(\Psi+2\Phi\right),\nonumber 
\end{eqnarray}
where an over-dot denotes the derivative with respect to the cosmic
time $t$ and the Hubble function is defined as $H\equiv\dot{a}/a$.
As can be easily recognized from Eq.~(\ref{eq:perX}), due to the
fact that in the quasi-static approximation $\hat{k}\gg1$ we do not
have any additional momentum scale in the theory, we cannot compensate
the contribution of the term $\hat{k}^{2}(\delta X+2\Psi+4\Phi)$;
hence, as a result from Eq.~(\ref{eq:perX}), we get 
\begin{equation}
\delta X=-2\left(\Psi+2\Phi\right).
\end{equation}
This outcome~\cite{Belgacem:2017ihm,Amendola:2017ovw} demonstrates
that, for the DW model in the quasi-static approximation, the perturbation
of the auxiliary field $X$ does not have a scale dependence. One
also obtains a similar result for the perturbed auxiliary field $U$
which is given by $\delta U=f'\delta X$. On the other hand, from
the perturbed spatial components of the Einstein equations for the
DW model, in the absence of the anisotropic stress, one obtains the
following relation between the Bardeen potentials and the perturbations
of the auxiliary fields 
\begin{equation}
\Psi+\Phi+f'\delta X+\delta U+\left(\Psi+\Phi\right)\left(f+U\right)=0.\label{eq:subhorij}
\end{equation}
By using Eq.~(\ref{eq:subhorij}) and the expressions for $\delta X$
and $\delta U$ we obtain the gravitational slip parameter~(\ref{slipDW})
for the DW model. So we once more confirm that $\eta$ for the DW
model in the quasi-static approximation is a scale-independent variable,
which in general has a value different from unity, and cannot be constrained
by the measurements of $c_{\text{T}}$.

The speed of gravitational waves within the DW model has been first
obtained in Ref.~\cite{Koivisto:2008dh}, where the author shows
that the gravitational waves in the DW model propagate with the speed
of light and have two orthogonal polarizations as in the case of GR.
Then, the DW model modifies only the gravitational waves amplitude
compared to GR, which is due to a different rate of the background
expansion in this model. This might lead to other observational effects
(see \cite{Belgacem:2017ihm,Amendola:2017ovw}).

\section{Shift symmetry: Covariant Galileon Theory}

We close our discussion with an example of the Galileon theory which
is given by the following action~\cite{Nicolis:2008in} 
\begin{equation}
S=\frac{1}{16\pi G}\int d^{4}x\sqrt{-g}\left[R+\frac{1}{2}\sum_{i=1}^{5}c_{i}\mathcal{L}_{i}\right]+S_{\text{m}},\label{galileonact}
\end{equation}
where $\left(c_{i},\left\lbrace i,1,5\right\rbrace \right)$ are the
free parameters of the model and the corresponding five covariant
Lagrangians $\mathcal{L}_{i}$ are defined as 
\begin{eqnarray}
\mathcal{L}_{1}=\tilde{M}^{3}\phi,\hspace{15mm}\mathcal{L}_{2}=\left(\nabla\phi\right)^{2},\hspace{15mm}\mathcal{L}_{3}=\left(\Box\phi\right)\left(\nabla\phi\right)^{2}/\tilde{M}^{3},\\
\mathcal{L}_{4}=\left(\nabla\phi\right)^{2}\left[2\left(\Box\phi\right)^{2}-2\phi_{;\mu\nu}\phi^{;\mu\nu}-R\left(\nabla\phi\right)^{2}/2\right]/\tilde{M}^{2},\\
\mathcal{L}_{5}=\left(\nabla\phi\right)^{2}\left[\left(\Box\Phi\right)^{3}-3\left(\Box\phi\right)\phi_{;\mu\nu}\phi^{;\mu\nu}+2\phi_{;\mu}^{;\nu}\phi_{;\nu}\phi^{;\rho}\phi_{;\rho}^{;\mu}-6\phi_{;\mu}\phi^{;\mu\nu}\phi^{;\rho}G_{\nu\rho}\right].
\end{eqnarray}
Here, the parameter $\tilde{M}$ should not be confused with a mass
term since is the coefficient of a term which is linear in the Galileon
field $\phi$.

The Galileon theory defined by the above mentioned action is invariant
under the field transformation 
\begin{equation}
\phi\rightarrow\phi+b_{\mu}x^{\mu}+C,
\end{equation}
where $b_{\mu}$ and $C$ are arbitrary constants. This symmetry is
also known as a Galileon symmetry. Being invariant under shift symmetry,
the Galileon theory does not introduce any additional mass scale into
the theory. Hence, in agreement with the above discussed examples,
the Galileon theory in the quasi-static approximation will have a
scale-independent gravitational slip parameter generally different
from unity. Indeed, in this case the linear perturbation equations
have the structure~\cite{DeFelice:2010as} 
\begin{eqnarray}
B_{6}\Phi+B_{7}\delta\phi+B_{8}\Psi=0,\label{perteqgalileon1}\\
D_{7}\Phi+D_{9}\delta\phi+D_{10}\Psi=0,\label{perteqgalileon2}
\end{eqnarray}
where $B_{i}$ and $D_{i}$ are scale-independent functions that depend
on $H$, $c_{i}$, $\tilde{M}$ and on the Galileon field $\phi$
and its time derivatives. Now, by using Eqs.~(\ref{perteqgalileon1})
and~(\ref{perteqgalileon2}) to subtract the perturbation of the
Galileon field $\delta\phi$, we obtain for the gravitational slip
parameter the following expression 
\begin{equation}
\eta=\frac{D_{10}B_{7}-D_{9}B_{8}}{D_{7}B_{7}-D_{9}B_{6}}.\label{etagalileon}
\end{equation}
So in agreement with the examples presented in the previous sections
we observe that for the Galileon theory in the quasi-static approximation
the gravitational slip parameter again does not have a scale dependence
and in general will not approach the GR value $\eta=1$. However,
as it has been shown in Ref.~\cite{Ezquiaga:2017ekz}, the gravitational
wave speed put stringent constraints on the structure of the functions
$B_{i}$ and $D_{i}$. Namely, it has been shown that in order to
pass the gravitational wave constraints the constants $c_{4}$ and
$c_{5}$ in the action~(\ref{galileonact}) must be vanishing. Under
these conditions the functions $B_{i}$ and $D_{i}$ in Eqs.~(\ref{perteqgalileon1})
and~(\ref{perteqgalileon2}) reduce to 
\begin{eqnarray}
B_{7}=0,\hspace{14.4mm}B_{6}=B_{8}=2M_{\text{pl}}^{2},\\
D_{9}=c_{2}-4c_{3}H\dot{\phi}/\tilde{M}^{3}-2c_{3}\ddot{\phi}/\tilde{M}^{3}.
\end{eqnarray}
Inserting these expressions into Eq.~(\ref{etagalileon}) we get
$\eta$ to be exactly unity. This case is therefore trivial, since
for $c_{4},c_{5}=0$ there is no modification of gravity at all.

\section{Conclusions}

The recent precise measurement of the GW speed/light speed ratio killed
several extended forms of modified gravity models, e.g. those in which
the coupling of a scalar field to gravity is not conformal. Even duly
remarking that the constraint applies only at the present time and
only to models in which both dark matter and baryons are coupled,
this historical observation has several important consequences. One
of these is that the observable anisotropy parameter $\eta$ has to
be unity at scales larger than the field mass scale. As noted in Ref.
\cite{Amendola:2017orw}, however, this applies only insofar the modification
of gravity introduces a new mass scale in the theory. In this paper
we point out that when no new mass scale is introduced (besides the
Planck mass) then \emph{a}) $\eta$ becomes scale-independent, \emph{b})
$\eta$ is in general different from unity, and \emph{c}) it is not
necessarily constrained by the GW speed.

To explore this scenario, we discussed three models in which no mass
scale arises: standard Brans-Dicke, nonlocal DW model, and Galileon
Lagrangians. In the first two cases $c_{\text{T}}$ is predicted to
be exactly unity, but $\eta$ is left unconstrained, scale-independent,
and in general different from unity. For the Galileon Lagrangian,
$\eta$ is indeed scale-independent, but the requirement $c_{\text{T}}=1$
imposes that the gravity sector is a standard one and therefore $\eta=1$.
Except for this case, our study confirms that $\eta$ remains a very
informative modified gravity parameter even after the GW events and
should be a prime target for observational studies moving forward.

\acknowledgments  We acknowledge support from DFG through
the project TRR33 ``The Dark Universe.''

\appendix

 \bibliographystyle{apsrev}
\bibliography{AnisoRefs,Henrik,Nelson}

\end{document}